\documentclass[reprint,superscriptaddress,amsmath,amssymb,aps,prl,floatfix]{revtex4-1}
\usepackage{color}
\usepackage{graphicx}
\usepackage{bm}

\newcommand\be{{\bm e}}
\newcommand\bv{{\bm v}}
\newcommand\bx{{\bm x}}
\newcommand\bF{{\bm F}}
\newcommand\beps{{\bm \varepsilon}}
\newcommand\bxi{{\bm \xi}}
\newcommand\bdeta{{\bm \eta}}
\newcommand\grad{{\bm \nabla}}

\newcommand\refSI{{(SM Text)}}
\newcommand\refSInb{{SM Text}}

\newcommand\refSMmovie{{Supplementary Video}}
\newcommand\refSMmovies{{Supplementary Videos}}

\begin{document}

\title{Autonomous actuation of zero modes in mechanical networks far from equilibrium}

\author{Francis G. Woodhouse}
\altaffiliation[Present address: ]{Mathematical Institute, University of Oxford, Andrew Wiles Building, Radcliffe Observatory Quarter, Woodstock Road, Oxford OX2 6GG, U.K.}
\email[Correspondence to: ]{francis.woodhouse@maths.ox.ac.uk}
\affiliation{ Department of Applied Mathematics and Theoretical Physics, Centre for Mathematical Sciences, University of Cambridge, Wilberforce Road, Cambridge CB3 0WA, U.K. }

\author{Henrik Ronellenfitsch}
\author{J\"orn Dunkel}
\affiliation{ Department of Mathematics, Massachusetts Institute of Technology, 77 Massachusetts Avenue, Cambridge MA 02139-4307, U.S.A. }

\date{\today}

\begin{abstract}
A zero mode, or floppy mode, is a non-trivial coupling of mechanical components yielding a degree of freedom with no resistance to deformation. Engineered zero modes have the potential to act as microscopic motors or memory devices, but this requires an internal actuation mechanism that can overcome unwanted fluctuations in other modes and the dissipation inherent in real systems. In this work, we show theoretically and experimentally that complex zero modes in mechanical networks can be selectively mobilized by non-equilibrium activity. We find that a correlated active bath actuates an infinitesimal zero mode while simultaneously suppressing fluctuations in higher modes compared to thermal fluctuations, which we experimentally mimic by high frequency shaking of a physical network. Furthermore, self-propulsive dynamics spontaneously mobilise finite mechanisms as exemplified by a self-propelled topological soliton. Non-equilibrium activity thus enables autonomous actuation of coordinated mechanisms engineered through network topology.
\end{abstract}

\maketitle

Soft, electronics-free assemblies capable of autonomous motion and reconfiguration are emerging as the basis of new adaptable smart materials.
Macroscopic morphology schemes, such as snap-through~\cite{2016Raney_PNAS,2017Gomez_NatPhys,2017Gomez_PRL,2017Chen,2015Overvelde_PNAS} and buckling~\cite{2018Fu_NatMat,2018Baek_PNAS,2014Kang_PRL} driven by heat~\cite{2017Ding_SciAdv} or chemo-fluidics~\cite{2016Wehner_Nature}, are complemented by the robustness of topological modes~\cite{2014Kane_NPhys,2015Lubensky_RepProgPhys,2018Zhou_PRL,2016Chen_PRL,2015Susstrunk_Science} to give a wide set of components based on elastic networks~\cite{2017Bertoldi_NatRevMat,2014Silverberg_Science,2017Overvelde_Nature,2017Yang_PNAS,2016Coulais_Nature}.
In such networks, a zero mode~(ZM) arises as a degree of freedom with no resistance to small deformation, either as an infinitesimal zero mode~(IZM) with resistance at nonlinear order~\cite{1991Calladine_IJSS,2006Heussinger_PRL,2017Coulais_Nature,2010Mao_PRL,2012Sun_PNAS,2003Guest_JMPS,2014Chen_PNAS,2015Paulose_NatPhys,2010Chen_PRL} or a mechanism with a continuous range of motion~\cite{2003Guest_JMPS,2017Rocklin_NatComms,2012Sun_PNAS,2009DeSouza_PNAS,2014Chen_PNAS}.
A designed ZM can potentially be exploited as a complex coupling~\cite{2017Rocks_PNAS,2017Yan_PNAS} in an internally-driven material.
However, actuation of a ZM can be hampered by indiscriminate simultaneous excitation of nonzero harmonic modes~(HMs), particularly in noisy microscopic systems~\cite{2015Dunn_Nature,2017Nava_PRL,2018Rocklin,2015Mao_NatComms,2016Zhang_PRE,2010Chen_PRL}.
Non-equilibrium processes~\cite{2016Shen_PRL}, which support intricate topological edge currents~\cite{2015Nash_PNAS,2015Wang_PRL,2017Souslov_NatPhys} and unorthodox stress responses~\cite{2016Ronceray_PNAS,2017Stam_PNAS}, may hold the key to overcoming this actuation dilemma.

In this work, we show that active matter provides effective schemes to autonomously actuate a mechanical ZM.
Active biophysical systems, such as bacterial suspensions or self-propelled microswimmers, convert disperse environmental energy into directed motion~\cite{2011Koch_ARFM,2013Marchetti_RMP,2016Bechinger_RMP}.
Tracers in an active bath, and the active particles themselves, then have positional statistics differing from thermal white noise~\cite{2010Rushkin_PRL,2014Maggi_PRL,2016Fodor_PRL}.
This statistical `colour', which depends on properties such as fuel availability and suspension density, can be used to drive mode actuation statistics away from equilibrium in a controllable fashion~\cite{2016Battle_Science,2011Franosch_Nature}, meaning features such as geometric asymmetry can be exploited to do work~\cite{2010DiLeonardo_PNAS,2010Sokolov_PNAS}.
First, we show that correlated noise generated by an active matter bath~\cite{2014Maggi_PRL} can actuate a complex mechanical IZM while markedly suppressing HMs to a degree dependent on temporal correlations, as well as exemplifying experimentally that fluctuation-based IZM actuation can be mimicked by simple high-frequency shaking.
We then broaden to self-propulsive Rayleigh activity~\cite{1998Schweitzer_PRL,2000Erdmann_EPJB,2017Forrow_PRL}, appropriate for a network whose nodes have intrinsic motility~\cite{2013Bricard_Nature}.
We show that this scheme mobilises a full mechanism comprising a propagating domain boundary in the SSH lattice~\cite{2014Chen_PNAS,2014Kane_NPhys}, suggesting that Goldstone modes of arbitrary complex systems can be mobilised by non-equilibrium driving~\cite{2016Battle_Science,2016Krishnamurthy_NatPhys,2017Vella_PNAS}.

\begin{figure}[b]
\includegraphics[width=\columnwidth]{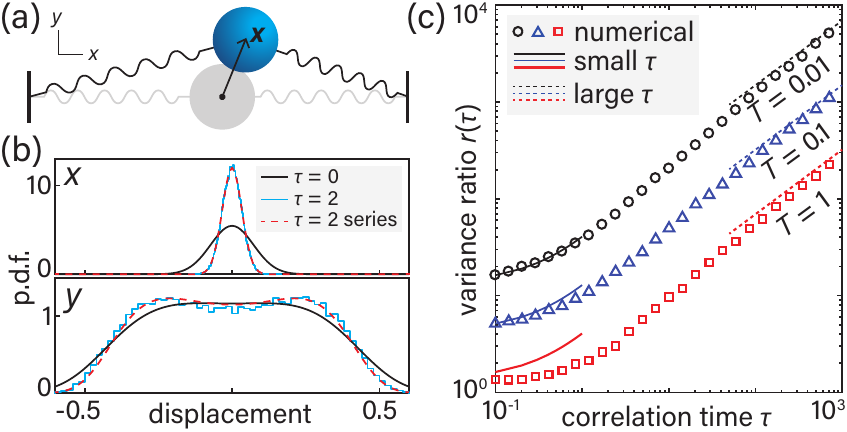}
\caption{Zero mode actuation in a bead--spring model.
(a)~Single mass at offset $(x,y)$, with HM in $x$ and IZM in $y$~\cite{2016Zhang_PRE}.
(b)~Comparison of $T = 0.01$ exact marginal densities of $x$ and $y$ at $\tau = 0$ (black) with $\tau = 2$ densities from simulation (blue) and approximated from Eqs.~\eqref{eq:onemass_H} and~\eqref{eq:onebead_Heff} (red dashed).
Histogram for $\tau = 2$ from $50000$ samples, others by quadrature \refSI.
(c)~Variance ratio $r(\tau) = \langle y^2 \rangle / \langle x^2 \rangle$ from simulations (markers) with approximations for small $\tau$ (solid lines) and large $\tau$ (dashed lines) from Eqs.~\eqref{eq:onebead_smallTau} and~\eqref{eq:onebead_largeTau}. $5000$ samples per marker; 95\% CIs smaller than markers \refSI.}
\label{fig:onebead}
\end{figure}

To gain intuition about the core idea, consider this basic example. A mass is held between two fixed points $\hat\bx = (1,0)$ and $-\hat\bx$ in the plane by two identical springs of unit natural length and stiffness (Fig.~\ref{fig:onebead}a)~\cite{2016Zhang_PRE}. Let $\bx = (x,y)$ be the offset of the mass from its equilibrium~$(0,0)$. The total elastic potential energy is
\begin{align}
H(\bx) = \tfrac{1}{2} \left[ (|\bx-\hat\bx|-1)^2 + (|\bx+\hat\bx|-1)^2 \right].
\label{eq:onemass_H}
\end{align}
If the mass is excited by thermal noise of temperature~$T$, its position is Boltzmann distributed with $p(\bx) \propto e^{-H/T}$.
To leading order in $T \ll 1$, $p$ can be approximated by $p_0(\bx) \propto e^{-H_0/T}$ where $H_0 = x^2 + \tfrac{1}{4}y^4$~\cite{2016Zhang_PRE}.
This bare Hamiltonian shows the two eigenmodes: an HM of frequency $\sqrt{2}$ parallel to the springs, and an IZM perpendicular to the springs. Scaling considerations then imply that the variances $\langle x^2 \rangle$ and $\langle y^2 \rangle$ vary as $T$ and $\sqrt{T}$, respectively.
The fine details can be seen by formal expansion in $\sqrt{T}$ \refSI, in which interaction cross-terms are negligible as $T \rightarrow 0$; this does not hold in the more complex examples below, where broken symmetries induce non-negligible interactions $\propto xy^2$ in $H_0$ that can cause strong violation of naive equipartition $\langle x^2 \rangle \sim T / \omega^2$.
Either way, the basic $T$-scalings still hold, so $\langle y^2 \rangle / \langle x^2 \rangle \rightarrow \infty$ as $T \rightarrow 0$. Thus fluctuations in the IZM dominate those in the HM at low temperature.

\begin{figure*}
\includegraphics[width=\textwidth]{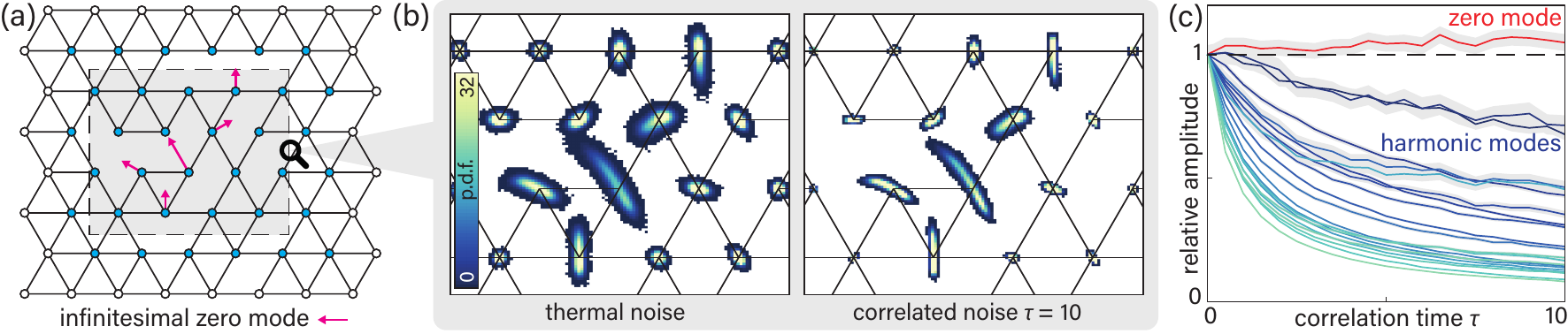}
\caption{Active noise actuates an IZM while suppressing HMs in a mechanical network (\refSMmovies~1 and~2).
(a)~A network of unit length, unit stiffness springs is designed to contain exactly one IZM (arrows). White nodes are pinned.
(b)~Histograms of node positions in highlighted area of (a) when actuated by thermal noise (left) and correlated noise with $\tau = 10$ (right) of strength $T = 10^{-3}$. The IZM is more cleanly actuated by correlated noise while HMs are suppressed.
(c)~Thermal-relative amplitude $\langle u_i^2 \rangle_\tau / \langle u_i^2 \rangle_{\tau = 0}$ of the 21 lowest-eigenvalue modes $u_i$ of (a) from numerical integration at $0 \leqslant \tau \leqslant 10$, with $T = 10^{-3}$ fixed to maintain constant intensity $\overline{ \langle v_{\alpha i}(0) v_{\beta j}(t) \rangle }$. The IZM (red) is barely affected, while HMs (blues) diminish. Grey areas are estimated 95\% CIs; data from $20$ realisations to $t = 2\times 10^4$ with $\delta t = 10^{-4}$ at each of $20$ values of $\tau$~\refSI.}
\label{fig:network}
\end{figure*}

Uncorrelated noise is a crude tool, with only one control parameter.
Actuation by biological or chemical active matter~\cite{2012Sanchez_Nature,2013Bricard_Nature} can allow finer tuning.
The forces generated by a motile bath can be modelled by an Ornstein--Uhlenbeck process  $\bxi$ obeying $\tau \dot\bxi = -\bxi + \bdeta$, where $\bdeta$ is a thermal process with variance $2\gamma T$ for amplitude $T$ and friction $\gamma$, and $\tau$ is the correlation time~\cite{2016Fodor_PRL,2018Sandford_PRE}.
These parameters depend on the properties of the active medium~\cite{2014Maggi_PRL}, and so by changing these properties---density, temperature, fuel concentration---the statistics can be tuned.
In the overdamped limit with $\gamma$ rescaled to one, it was shown in Ref.~\cite{2016Fodor_PRL} that small $\tau \ll 1$ adjusts $H$ to an effective potential
\begin{align}
    H_\text{eff} = H + \tau \left[ \tfrac{1}{2}|\grad H|^2 - T\nabla^2 H \right] + O(\tau^2).
\label{eq:onebead_Heff}
\end{align}
For our example in Fig.~\ref{fig:onebead}, this (or the unified coloured noise approximation~\cite{1987Jung_PRA,2015Maggi_SciRep}) gives a low-$T$ expansion $H_\text{eff} = \left(1 + 2\tau \right) x^2 + \frac{1}{4} y^4   + \cdots$, implying
\begin{align}
\frac{\langle y^2 \rangle}{\langle x^2 \rangle} \approx \frac{4\Gamma(\tfrac{3}{4})}{\Gamma(\tfrac{1}{4})}\frac{1+2\tau}{\sqrt{T}}
\label{eq:onebead_smallTau}
\end{align}
for small $T$ and $\tau$ (\refSInb; Fig.~\ref{fig:onebead}c).
This is reflected in the $\tau = 2$ marginal densities (Fig.~\ref{fig:onebead}b): $x$ contracts, while $y$ is unchanged except for a small bimodality directly analogous to boundary accumulation of microswimmers~\cite{2018Das_NJP}.
In the opposite limit $\tau \gg 1$, where Eq.~\eqref{eq:onebead_Heff} does not hold, we show in the \refSInb~that separation of time scales gives the asymptotic power law
\begin{align}
    \frac{\langle y^2 \rangle}{\langle x^2 \rangle} \approx \frac{ 2^{7/3} \Gamma(\tfrac{5}{6})}{\pi^{1/2}} \left(\frac{\tau}{T}\right)^{2/3}
    \label{eq:onebead_largeTau}
\end{align}
as $\tau/T$ grows large (Fig.~\ref{fig:onebead}c).
These results demonstrate that active noise is an effective means to actuate an IZM.

\begin{figure}[t]
\includegraphics[width=\columnwidth]{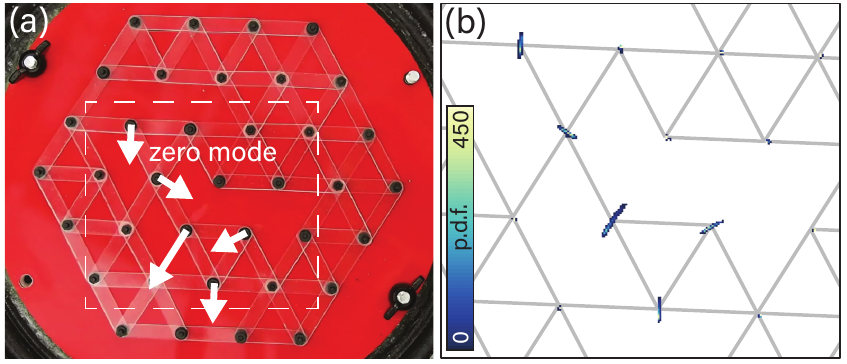}
\caption{Shaking actuates the IZM of a stiff mechanical network (\refSMmovie~3).
(a)~Photograph of network with isolated IZM (arrows) constructed from plastic. Joints are freely-rotating pins, with boundary nodes fixed in position.
(b)~Actuating the network by high frequency shaking mobilises the IZM, shown by node positional histograms for the subregion indicated in (a) computed by particle tracking over $14\,\text{min}$ with distances rescaled by mean edge length.
(See \refSInb~for experimental methods.)
}
\label{fig:experiment}
\end{figure}

\begin{figure}[b]
\includegraphics[width=\columnwidth]{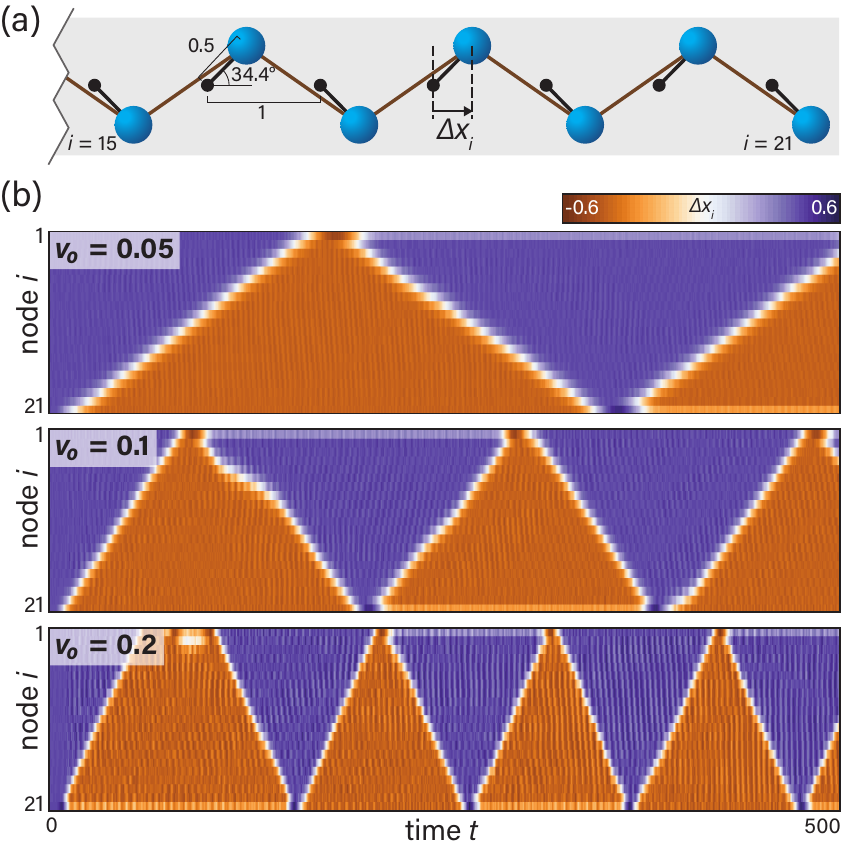}
\caption{Self-propulsive activity spontaneously mobilises a complex mechanism (\refSMmovie~5).
(a)~Mechanical SSH model~\cite{2014Chen_PNAS,2014Kane_NPhys}, which has a periodic mechanism comprising progressive flipping of the masses from right to left and back again. Black nodes are fixed, blue nodes are mobile.
(b)~Endowing a $21$-node chain with self-propulsive activity spontaneously mobilises the mechanism. The progression speed depends on the propulsion $v_0$, seen through the time-dependent offsets $\Delta x_i$ of mobile nodes from their pinning points.
All bonds of strength $k=10$ with $\gamma_0 = 1$, integrated at $\delta t = 10^{-5}$ from initial random perturbation.
\label{fig:SSH}}
\end{figure}

In a general mechanical network, the same ideas can be used to actuate an IZM with many masses moving in a complex, coordinated fashion.
A stable mechanical network comprises nodes $\alpha$ with rest positions $\bx_\alpha^0$, and Hookean elastic bonds $(\alpha,\beta)$ between nodes $\alpha$ and $\beta$ of natural lengths $\ell_{\alpha\beta} = |\bx_\alpha^0 - \bx_\beta^0|$ and stiffnesses $k_{\alpha \beta} > 0$.
When the nodes are perturbed to $\bx_\alpha$ the network has elastic energy
\begin{align*}
H(\{\bx_\alpha\}) = \frac{1}{2} \sum_{(\alpha,\beta)} k_{\alpha\beta} (|\bx_\alpha - \bx_\beta| - \ell_{\alpha\beta})^2,
\end{align*}
which is minimised if (but not always only if) $\bx_\alpha = \bx_\alpha^0$. Node $\alpha$ then feels an elastic force $\bF_\alpha = -\grad_\alpha H$, where we notate $\grad_\alpha \equiv \partial/\partial \bx_\alpha$.
Large real-world systems also possess dissipation, which we include as a linear friction force $\gamma\dot\bx_\alpha$. In general dissipation within bonds~\cite{2016Raney_PNAS} could also be included as forces $\propto \dot\bx_\alpha - \dot\bx_\beta$ giving a non-diagonal friction matrix~\cite{RayleighVol2}.
We further restrict to two dimensions, and assume all nodes are identical, adopting units in which node masses are one and other constants are scaled by a characteristic stiffness and length~\refSI.
Finally, we pin boundary nodes (Figs.~\ref{fig:network} and \ref{fig:experiment}) or particular interior nodes (Fig.~\ref{fig:SSH}) to eliminate rigid body translations and rotation.

Without any actuation, the mobile nodes obey passive force balance, namely $\ddot \bx_\alpha = -\grad_\alpha H - \gamma \dot \bx_\alpha$.
Consider a small perturbation $\bx_\alpha = \bx_\alpha^0 + \beps_\alpha$ of the rest state, and let~$\be$ be the vector obtained by flattening $\beps_\alpha$. To first order, $\be$ obeys $\ddot e_i = -\sum_j D_{ij} e_j - \gamma \dot e_i$, where the Hessian~$D_{ij} = \partial_i \partial_j H(\bx^0)$ is the dynamical matrix~\cite{2014Kane_NPhys}.
The orthonormal eigenvectors $\{\bv_k\}$ of $D_{ij}$ give the fundamental modes of the elastic network, whose non-negative eigenvalues $\{ \omega_k^2 \}$ determine if each mode is an HM ($\omega_k^2 > 0$) or a ZM ($\omega_k^2 = 0$). These then form a basis for configurations $\bx_\alpha$. Writing $c_k(t)$ for the component of mode $k$ at time $t$, given by dotting the flattened $\bx_\alpha(t)$ with $\bv_k$, we define the amplitude of mode~$k$ to be the time average of its squared coefficient,~$\langle c_k(t)^2 \rangle$.
Our goal is to show that the amplitudes of ZMs can be selectively actuated in suitably designed networks.

When a network is actuated by active forces on its nodes, its dynamics depends on both its structure and the type of activity.
The positions $\bx_\alpha$ obey the general actuated dynamics
\begin{align}
\ddot \bx_\alpha = -\grad_\alpha H - \gamma \dot \bx_\alpha + \bF_\alpha(\dot\bx_\alpha; t),
\label{eq:dyn_basic}
\end{align}
where $\bF_\alpha$ represents the actuation process. For this process we will consider not only the thermal and correlated active bath processes exemplified above, but also strong self-propulsive activity through internal energy depot actuation. These incorporate increasing levels of non-equilibrium dynamics which progressively actuate infinitesimal and finite zero modes.

Generalising the single mass example above, a mechanical network driven by Ornstein--Uhlenbeck noise can be formulated with an extra vector $\bxi_\alpha$ for each node~\cite{2016Fodor_PRL,2017Sandford_PRE,2018Das_NJP}. These follow
\begin{align}
    \tau  \dot\bxi_\alpha = -\bxi_\alpha + \bdeta_\alpha
\label{eq:net_OUnoise}
\end{align}
for independent Gaussian noise processes $\{\eta_{\alpha i}\}$ of variances $2\gamma T$.
We set $\bF_\alpha = \bxi_\alpha$, giving actuation forces correlated as $\langle \xi_{\alpha i}(t) \xi_{\beta j}(t') \rangle =  \delta_{\alpha\beta}\delta_{ij} (\gamma T/\tau) e^{-|t-t'|/\tau}$. The limit $\tau \rightarrow 0$ then gives thermal noise.
We also again take the overdamped limit, appropriate for immersion in an active matter bath. After rescaling to set $\gamma$ to $1$, this gives first-order dynamics
\begin{align}
\dot \bx_\alpha = -\grad_\alpha H + \bxi_\alpha
\label{eq:net_corrdyn}
\end{align}
subject to Eq.~\eqref{eq:net_OUnoise}.
Varying $\tau$ while holding $T$ fixed then probes the effect of increasing activity-driven correlations at constant actuation intensity $\overline{ \langle \xi_{\alpha i}(0) \xi_{\beta j}(t) \rangle }$.

For a mechanical network designed to have a non-trivial isolated IZM (Fig.~\ref{fig:network}a; \refSInb), correlated noise highlights the IZM and suppresses  fluctuations at other nodes. Numerical integration of Eq.~\eqref{eq:net_corrdyn} shows that, while thermal noise ($\tau = 0$) actuates the IZM with significant surrounding fluctuations (Fig.~\ref{fig:network}b; \refSMmovie~1), an active process with $\tau > 0$ damps HM fluctuations relative to those of the IZM (Fig.~\ref{fig:network}b; \refSMmovie~2).
This is confirmed in the mode amplitudes $\langle c_k^2 \rangle$, shown in Fig.~\ref{fig:network}c: the IZM amplitude remains at its $\tau = 0$ level, while HMs decay as $\tau$ increases.
Two further examples showing IZM preservation and HM suppression are given in the \refSInb.
The same suppression is not guaranteed if the ZM contains a low-coordination node with bistability in its position~\refSI, since the fluctuation basis depends on the state of the bistable node, but this can typically be avoided in design.

Even without correlation, persistent low-temperature ZM--HM coupling enhances mode amplitudes and causes naive equipartition to fail. In a system comprising only HMs $u_i$ of frequencies $\omega_i$, two-mode interactions are diagonalised away and the lowest-order terms remaining are at best third order. For low temperatures $T \ll 1$ interactions can thus be neglected relative to harmonic energies $u_i^2$ and simple equipartition of independent modes, $\langle u_i^2 \rangle \approx T/\omega_i^2$, is a good amplitude estimate.
However, with a quartic ZM $v$ present, there are three types of lowest-order term in $T \ll 1$: the independent mode energies $v^4$ and $u_i^2$, and ZM--HM interactions $u_i v^2$.
Interactions contribute an effective repulsive quartic potential on the ZM, causing its amplitude $\langle v^2 \rangle$ to increase with every additional HM interaction \refSI.
Moreover, the HM amplitudes $\langle u_i^2 \rangle$ are also increased by their interactions with the ZM \refSI.

Small- and large-$\tau$ asymptotics provide general principles for the behaviour of network modes.
When $\tau \ll 1$, computing the effective potential in Eq.~\eqref{eq:onebead_Heff}~\cite{2016Fodor_PRL} for an arbitrary small-$T$ expansion $H \approx \sum_i a_i u_i^2 + \sum_i b_i u_i v^2 + A v^4$ in HMs $u_i$ and a quartic IZM~$v$ gives new effective coefficients incorporating $\tau$ to leading order~\refSI. In particular, the stiffnesses $a_i$ become $a_i(1 + 2\tau a_i)$, showing that weaker modes have a proportionally weaker response to increasing $\tau$, while $\langle v^2 \rangle$ is unchanged.
Conversely, as $\tau \rightarrow \infty$, a scaling analysis shows that an IZM has amplitude proportional to $(T/\tau)^{1/3}$ while all HMs have amplitude proportional to $(T/\tau)$~\refSI, generalising our earlier example.
Physical dependencies are clarified by restoring dimensionful parameters, turning $T/\tau$ into $\gamma T / \kappa^2 \lambda^2 \tau$ for typical spring stiffness $\kappa$ and length $\lambda$.

In a real mechanical network, local activity can be mimicked by the randomisation generated through high frequency shaking.
We fabricated a stiff-edged network designed to contain an isolated IZM (Fig.~\ref{fig:experiment}a) by connecting plastic edges with freely-rotating joints.
We then actuated the network by mounting it on a baseboard, with fixed edge nodes, and securing this board to a loudspeaker~\refSI.
The speaker was driven at $49\,\text{Hz}$ with $20\%$ Ornstein--Uhlenbeck noise of correlation time $(1/345)\,\text{s}$ to prevent metastable sticking~\refSI.
The baseboard collisions generated by the high-frequency shaking randomises the motion of the IZM, actuating it through motion allowed by slight pliability in the pin joints (Fig.~\ref{fig:experiment}b and \refSMmovie~3).
The effective activity, and so the relative actuation of the IZM and HMs, can be controlled by changing the actuation frequency~\refSI.
Thus even elementary actuation strategies can be of practical use for IZM mobilisation.

We now turn to a stronger form of activity.
If the masses themselves are motile, able to convert chemical energy to kinetic energy, an effective model is to introduce what amounts to a negative frictional response at low speeds~\cite{2000Erdmann_EPJB,2017Forrow_PRL}.
Appealing to expansion techniques, as in the Toner--Tu model~\cite{1998Toner_PRE}, we use simple Rayleigh activity~\cite{RayleighVol2} with a velocity-dependent propulsive force
\begin{align}
\bF_\alpha = \gamma_f(1 - |\dot \bx_\alpha|^2/v^2)\dot\bx_\alpha,
\label{eq:rayleigh}
\end{align}
where $\gamma_f$ sets the force strength and $v$ is a natural speed in the absence of friction.
The overall effective friction force can then be viewed as $\gamma \dot\bx_\alpha - \bF_\alpha \equiv f(|\dot\bx_\alpha|)\dot\bx_\alpha$ with effective friction coefficient $f(u) = \gamma - \gamma_f + \gamma_f u^2 / v^2$.
Provided $\gamma_0 \equiv \gamma - \gamma_f < 0$, this gives a natural speed $v_0 = v|\gamma_0/\gamma_f|^{1/2}$ below which $f(u) < 0$. The quiescent state is then rendered unstable: to linear order, a perturbation of any eigenmode grows at least as fast as $e^{|\gamma_0/2| t}$ \refSI. Nonlinear effects bound this growth, giving oscillatory trajectories for constrained systems~\cite{2000Erdmann_EPJB}.

Under Rayleigh actuation, deviations in the ZM dominate those in the HM at small~$v_0$, as for correlated noise, with extra oscillatory structure in time.
Take once more the example in Fig.~\ref{fig:onebead}. At small $\gamma_0$, the leading order behaviour is a conservative oscillator $\ddot\bx = -\grad H$ whose amplitude is set by kinetic-to-potential energy balance $v_0^2 \sim H$ \refSI. The ZM--HM variance ratio therefore scales as $\langle y^2 \rangle / \langle x^2 \rangle \sim 1/v_0$ which diverges as $v_0 \rightarrow 0$.
A similar divergence occurs in the `overdamped' case $\gamma_0 \gg 1$.
This same suppression of HMs persists in more complex networks (\refSInb; \refSMmovie~4).

Beyond IZMs, self propulsion distinguishes itself from an active bath in its ability to fully mobilise free-moving mechanisms~\cite{2018Rocklin} even in topologically complex cases.
For a node tethered to just one fixed point, invariance means that the dynamics of the rotation angle $\theta$ has no elastic term. The node therefore accelerates and sustains an angular velocity $\sim v_0$ in the finite mechanism.
This principle generalises to intricate mechanisms of large mechanical networks.
Recently, a mechanical chain inspired by the SSH model of polyacetylene (Fig.~\ref{fig:SSH}a) has emerged as a rich source of topological phenomena~\cite{2014Chen_PNAS,2014Kane_NPhys}. This chain possesses a ZM localised at the boundary~\cite{2014Kane_NPhys} which gives rise to a finite mechanism manifesting as a domain boundary propagating along the chain~\cite{2014Chen_PNAS}.
Ordinarily, to begin propagating an external energy input is needed, either by a manual `kick' of the end nodes or by a global force field.
But persistent propagation is difficult: motion by a `kick' will inevitably slow and stop due to dissipation, while an external field needs regular adjustment to keep the defect moving between the two ends of the finite chain.
Endowing each node with active propulsion spontaneously actuates the ZM, mobilising the boundary and causing a domain wall soliton to propagate autonomously along the chain (Fig.~\ref{fig:SSH}b; \refSMmovie~5). While under passive dynamics the propagation speed of the topological soliton is set by the initial kinetic energy or the external force strength, here this is controlled by $v_0$, the effective self-propulsive speed (Fig.~\ref{fig:SSH}b).
The defect can propagate cleanly for multiple cycles up and down the chain, with occasional stalls or reversals caused by the complex interactions of the fluctuating active nodes~\refSI.

To conclude, we have shown that both non-equilibrium active baths and intrinsic self-propulsion can actuate infinitesimal and finite zero modes of mechanical networks.
Rapid progress in developing artificial active systems~\cite{2012Sanchez_Nature,2013Bricard_Nature} suggests practical routes to engineer active mechanical networks exhibiting fine-tuned fluctuation spectra or even complex response phenomena, such as non-reciprocity~\cite{2014Fleury_Science,2017Coulais_Nature} enabled by asymmetric nonlinear zero modes and non-equilibrium steady state statistics.
These networks could be used to perform complex mechanical tasks, such as enhancing transport of transiently-bound colloids, or to extract work from an active medium~\cite{2016Krishnamurthy_NatPhys} by attaching magnetic beads to drive a miniature dynamo, for example.
In general, we expect that any Goldstone mode of a complex mechanical system can be selectively mobilised by non-equilibrium activity of this kind~\cite{2016Battle_Science}.

\begin{acknowledgments}
We thank Pedro S\'{a}enz for assistance with the experimental setup, and Ellen A.\ Donnelly and Martin Zwierlein for helpful discussions.
This work was supported by Trinity College, Cambridge (F.G.W.), a GPU donation from NVIDIA Corporation (F.G.W.), an Edmund F. Kelly Research Award (J.D.), and a James S. McDonnell Foundation Complex Systems Scholar Award (J.D.).
\end{acknowledgments}

\end{document}